\documentclass[prl, twocolumn,superscriptaddress,nobibnotes]{revtex4-2}
\usepackage{graphicx}
\usepackage{amssymb}
\usepackage{amsmath}
\usepackage{bm}
\usepackage{color}
\usepackage{float}
\usepackage{setspace}
\usepackage{physics}
\usepackage{subfigure}
\usepackage{lineno}

\large

\begin{document}
	\title{Electrically tunable nonrigid moir\'e exciton polariton supersolids at room temperature} 
	
	\author{Xiaokun Zhai}
	\affiliation{Department of Physics, School of Science, Tianjin University, Tianjin 300072, China} 
	
	\author{Junhui Cao}
	\affiliation{Abrikosov Center for Theoretical Physics, Moscow Center for Advanced Studies, Kulakova str. 20, Moscow, 141701, Russia} 
	
	\author{Chunzi Xing}
	\affiliation{Department of Physics, School of Science, Tianjin University, Tianjin 300072, China}
	
	\author{Xinmiao Yang}
	\affiliation{Department of Physics, School of Science, Tianjin University, Tianjin 300072, China}
	
	\author{Xinzheng Zhang}
	\affiliation{The MOE Key Laboratory of Weak-Light Nonlinear Photonics and International Sino-Slovenian Join Research Center on Liquid Crystal Photonics, TEDA Institute of Applied Physics and School of Physics, Nankai University, Tianjin 300457, China}
	
	\author{Haitao Dai}
	\affiliation{Department of Physics, School of Science, Tianjin University, Tianjin 300072, China}
	
	\author{Xiao Wang}
	\affiliation{College of Materials Science and Engineering, Hunan University, Changsha 410082, China}
	
	\author{Anlian Pan}
	\affiliation{College of Materials Science and Engineering, Hunan University, Changsha 410082, China}

	\author{Stefan Schumacher}
	\affiliation{Department of Physics and Center for Optoelectronics and Photonics Paderborn (CeOPP), Universit\"{a}t Paderborn, 33098 Paderborn, Germany}
	\affiliation{Institute for Photonic Quantum Systems (PhoQS),
		Paderborn University, 33098 Paderborn, Germany}
	\affiliation{Wyant College of Optical Sciences, University of Arizona, Tucson, AZ 85721, USA}
	
	\author{Alexey Kavokin}
	\affiliation{Abrikosov Center for Theoretical Physics, Moscow Center for Advanced Studies, Kulakova str. 20, Moscow, 141701, Russia}
	\affiliation{School of Science, Westlake University, 18 Shilongshan Road, Hangzhou 310024, Zhejiang Province, China}
	
	\author{Xuekai Ma}
	\affiliation{Department of Physics and Center for Optoelectronics and Photonics Paderborn (CeOPP), Universit\"{a}t Paderborn, 33098 Paderborn, Germany}
	
	\author{Tingge Gao}
	\affiliation{Department of Physics, School of Science, Tianjin University, Tianjin 300072, China} 
	
	\begin{abstract}
		
		A supersolid is a macroscopic quantum state which sustains superfluid and crystallizing structure together after breaking the U(1) symmetry and translational symmetry. On the other hand, a moir\'e pattern can form by superimposing two periodic structures along a particular direction. Up to now, supersolids and moir\'e states are disconnected from each other. In this work we show that exciton polariton supersolids can form moir\'e states in a double degenerate parametric scattering process which creates two constituent supersolids with different periods in a liquid crystal microcavity. In addition, we demonstrate the nonrigidity of the moir\'e exciton polariton supersolids by electrically tuning the wavevector and period of one supersolid component with another one being fixed. Our work finds a simple way to link moir\'e states and out-of-equilibrium supersolids, offering to study nontrivial physics emerging from the combination of moir\'e lattices and supersolids which can be electrically tuned at room temperature in the future.

	\end{abstract}
	
	\maketitle
	

	Bose Einstein condensates, superfluids and other macroscopic quantum systems like superconductivity are coherent states which enrich our understanding of quantum physics \cite{cold atom BEC, BEC superfluid}. Recently, supersolids \cite{supersolid proposal 1, supersolid proposal 2, supersolid proposal 3, supersolid proposal 4} have attracted intensive attention and been investigated firstly in the liquid helium \cite{helium}, then experimentally realized in the cold polar and spin-orbit coupled quantum gases \cite{supersolid dipolar 1, supersolid dipolar 2, supersolid dipolar 3, supersolid dipolar 4, supersolid dipolar 5, supersolid SOI 1, supersolid SOI 2} and Bose Einstein condensates coupled to two optical cavities \cite{supersolid cavity 1, supersolid cavity 2}. This exotic phase exists in the superfluid with global phase coherence and crystallizing structure, after the onset of the broken U(1) symmetry and translational symmetry. An important question related to the supersolid is the stiffness or rigidity of the fringes, which can be explored by collective excitation, such as Goldstone modes and Higgs modes with particular relations to the order parameters \cite{supersolid dipolar 5, supersolid cavity 1}. Especially the spin-orbit coupled supersolid stripe patterns are found to be nonrigid theoretically \cite{nonrigid 1} and experimentally \cite{nonrigid 2} where the Raman coupling can tune the fringe's period and orientation and induce oscillating dynamics of the crystallizing mode. 
	
	
	Above supersolid phases are limited at ultralow temperature for the Bose Einstein condensates and superfluids of the cold quantum atom gases. On the other hand, at relatively higher temperature (4 K), the supersolid of the composite bosonic quasiparticles, such as exciton polaritons \cite{microcavity book} formed due to the strong coupling between excitons and photonic bound states in the continuum (BIC), within a microcavity with a subwavelength grating structure is also realized \cite{polariton supersolid1, polariton supersolid2}.The formation of the polariton supersolid is obtained by the $\chi$$^3$ nonlinearity induced degenerate optical parametric scattering \cite{OPO1,OPO2,OPO3,PRL1,Nanophotonics1} between two branches within the grating structure. By increasing the pumping density to manipulate the wavevectors of the density modulation mode through the polariton-polariton interaction in the supersolid phase, the tuning of the rigidity of the density pattern is usually below the resolution limit of the optical spectroscopy setup and thus difficult to be measured in experiments. In this context, alternative methods to study the nonrigidity of exciton polariton supersolids are expected.

	Based on the liquid crystal (LC) molecules inserted as the cavity spacer layer \cite{1-liquid crystal_science, stripe phase}, a new method to electrically tune exciton polaritons is realized \cite{Yao RD, Xiaokun vortex, gao ying}. In the LC microcavity, the applied electric field can control the orientation of the LC molecule director, thus the effective refractive index of the cavity is tuned and the cavity photon modes or polariton modes can be manipulated electrically (Figure 1(a)). More importantly, polariton condensates can also be tuned electrically with the energy modulation much larger than the blueshift caused by increasing the pumping density. In this regard, exciton polaritons in the LC microcavity offer a perfect platform to study the nonrigidity of the supersolid at room temperature. 
	
	\begin{figure}
		\centering
		\includegraphics[width=1\linewidth]{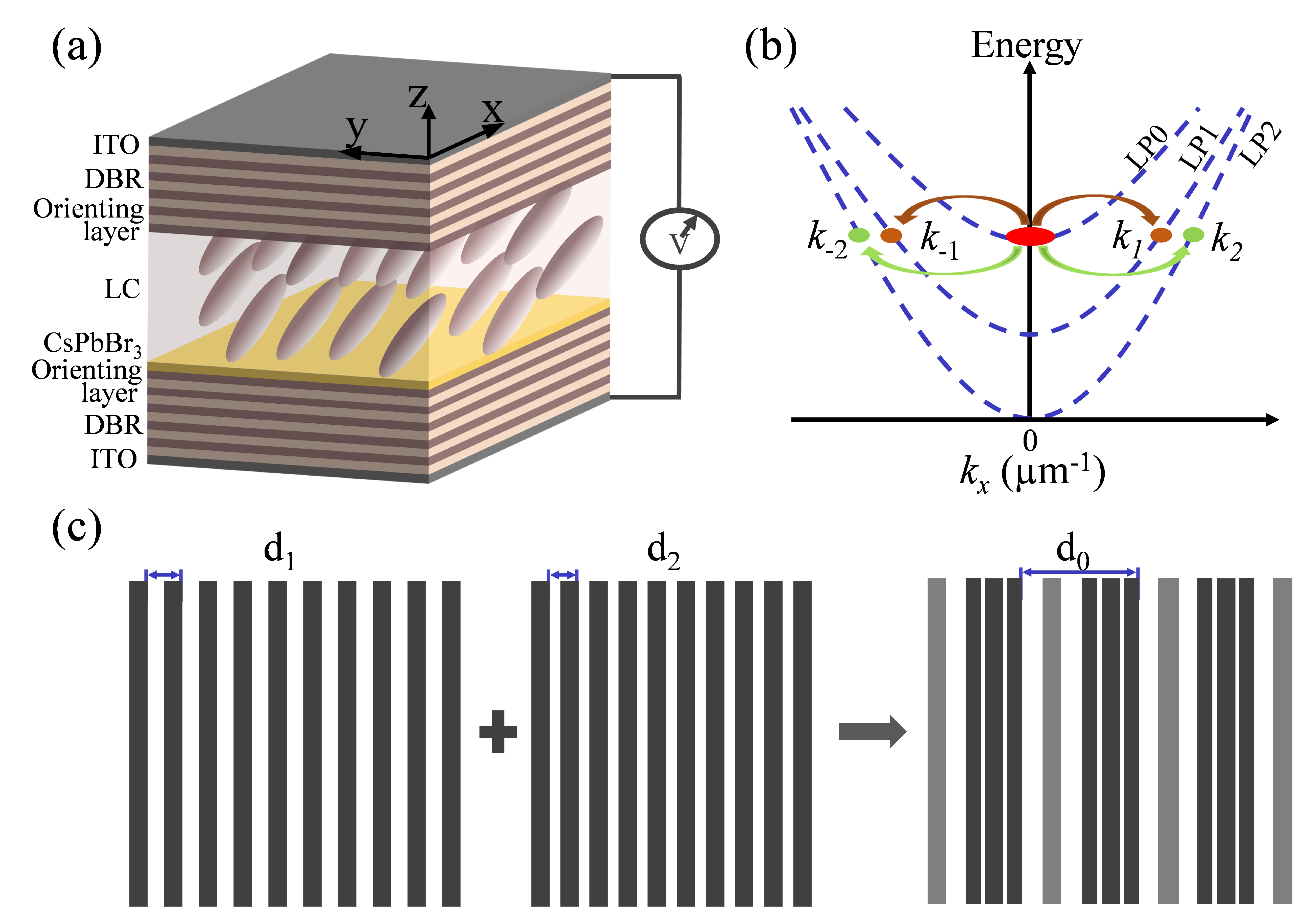}
		\caption{\textbf{{Schematic graph of the LC microcavity and degenerate parametric scattering.}} (a) The structure of the LC microcavity with \textit{x, y, z} axes indicated. (b) The double degenerate parametric scattering between LP0, LP1, and LP2. LP0, LP1 and LP2 are schematics of the polariton modes in the microcavity. (c) Illustration of the moir\'e exciton polariton supersolid formed by two supersolids in the degenerate parametric scattering process with different periods \textit{d1} and \textit{d2}. }
	\end{figure}

	In this work, we show that two spontaneously formed exciton polariton supersolids can form an exotic moir\'e state in a LC microcavity without the need of two constituent layers or structures. We also reveal the nonrigidity of this new kind of quantum state, where the moir\'e exciton polariton supersolids can be tuned electrically, thanks to the LC molecule within the microcavity. Our microcavity uses perovskite CsPbBr$_3$ microplates as the gain material, which has a large exciton binding energy and oscillator strength (Figure 1(a)) \cite{21-xiong condensation, zhuxiaoyang}. The appearance of the moir\'e exciton polariton supersolid within our microcavity is due to the coexistence of two degenerate parametric scattering processes from the ground state of one branch to the nonzero wavevector states of another two adjacent branches, as shown in the schematic graphs in Figure 1(b) (these modes are the schematics of the polaritons in the microcavity) and Figure 1(c). By modulating the nonzero wavevectors of the states within one branch, the moir\'e exciton polariton supersolid can be electrically manipulated into different moir\'e patterns, clearly confirming the nonrigidity of the supersolid. Our work links moir\'e physics and out-of-equilibrium supersolids together within a quantum fluid of light at room temperature and offers to study moir\'e induced nontrivial physics in the supersolid phase, which can be electrically tuned in the future. 
	
	In the experiments, the CsPbBr$_3$ microplate with the length of 70 $\mu$m and width of 80 $\mu$m is fabricated using the Chemical Vapor Deposition (CVD) method. A layer of SiO$_2$ is deposited onto the top of the CsPbBr$_3$ microplate to protect against water or other degradation factors. The ordering layer is spin coated afterwards and the microcavity is pasted with the same procedure as \cite{Yao RD}. Finally, the nematic liquid crystal molecule E7 is introduced into the microcavity (The details and structure of the microcavity are discussed and shown in the SM).
	
	The energy-\textit{k$_x$} dispersion of the LC microcavity is measured by using a home-made angle-resolved spectroscopy (shown in the SM) under the excitation of a femtosecond laser (spot size: 50 $\mu$m; repetition rate: 1000 Hz; wavelength: 400 nm; pulse width: 50 fs). Figure 2(a) shows the horizontally linearly polarized dispersions, we observe multiple lower-branch exciton polariton dispersions (LP0-LP2 are indicated) with different detunings (energy difference between the cavity photon mode and exciton resonance at normal incidence) due to the long cavity length of the LC microcavity, where these cavity modes strongly couple with the excitons in the perovskite CsPbBr$_3$ microplates. Note that the dispersions become flat at large-wavevector states, which confirms the existence of strong coupling within the LC microcavity (a dispersion measured with larger NA is provided in the SM). On the other hand, Figure 2(b) plots the vertical linearly polarized dispersion of the polariton modes (energy difference comes from the anisotropy of the LC molecule).

	By increasing the pumping density of the femtosecond laser, we find polaritons condense at the ground state $\ket{k_0}$ of LP0 \cite{nonequilibrium}, which can be seen from the dispersion measured at around 10.5 $\mu$J/cm$^2$ (plotted in Figure 2(c)). Under this pumping density, the integrated intensity of the emitted photons from the LC microcavity increases superlinearly, where the linewidth of the modes is reduced from 4 meV to 1.5 meV (Figure 2(d)) limited by the resolution of our spectrometer, showing a clear threshold behavior for polariton condensation below the Mott density. 
	
	Further enhancing the pumping density, polaritons begin to macroscopically occupy the states $\ket{\pm{k_1}}$ with the wavevector of $\pm$4.09 $\mu$m$^{-1}$ of LP1 and the states $\ket{\pm{k_2}}$ with the wavevector of $\pm$6.06 $\mu$m$^{-1}$ of LP2, as plotted in Figure 2(e). The integrated intensity changes of the modes $\ket{+k_1}$ against the pumping density are shown in Figure 2(f). We can clearly see the population of above state within LP1 increases with a threshold larger than the state $\ket{k_0}$ of LP0 as the function of the pumping density, due to the nonlinear stimulated scattering process among the LP0 and LP1 under the pumping density of 11.5 $\mu$J/cm$^2$. Similar integrated intensity and linewidth changes with the pumping density for the modes $\ket{+k_2}$ within LP2 are shown in the SM, where the coherence is confirmed for $\ket{+k_1}$ and $\ket{+k_2}$ state. This demonstrates the existence of two degenerate parametric scattering processes within the LC microcavity under higher pumping densities. That is, one occurs between the ground state $\ket{k_0}$ of LP0 and the states $\ket{\pm{k_1}}$ of LP1($k_0$=0, $k_1$=4.09 $\mu$m$^{-1}$), while another occurs between the ground state $\ket{k_0}$ of LP0 and the modes $\ket{\pm{k_2}}$ of LP2 ($k_0$=0, $k_2$=6.06 $\mu$m$^{-1}$). The parametric scattering occurs along the direction related with the anisotropy of the microcavity \cite{Xiaokun vortex}. We note that the parametric scattering between the polariton branches with opposite linear polarizations can occur when appropriate conservation constraint is obtained \cite{polarization OPO}.  
	
	\begin{figure}
		\centering
		\includegraphics[width=1\linewidth]{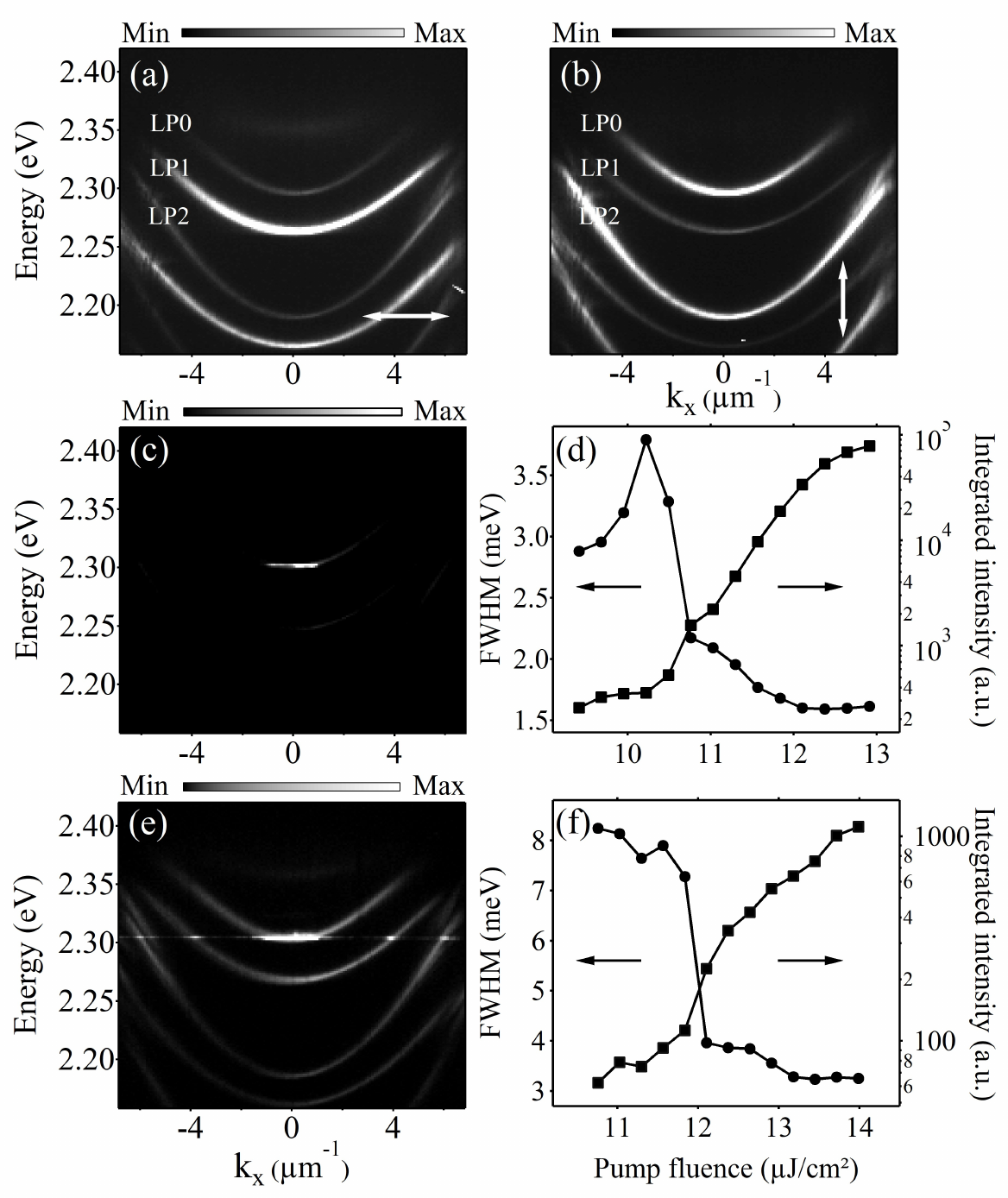}
		\caption{\textbf{{Degenerate parametric scattering within the microcavity.}} (a, b) Horizontally (\textit{x}) and vertically (\textit{y}) linearly polarized dispersions below threshold. The arrows indicate the linear polarization directions. (c, e) Total polariton dispersion above threshold under the pumping density of 10.5 $\mu$J/cm$^2$ and 11.5 $\mu$J/cm$^2$. Colorbar is in the linear scale. (d, f) Integrated intensity and linewidth of polariton modes at the ground state $\ket{k_0}$ of LP0 and $\ket{+k_1}$ of LP1.}
	\end{figure}

	\begin{figure}
		\centering
		\includegraphics[width=1\linewidth]{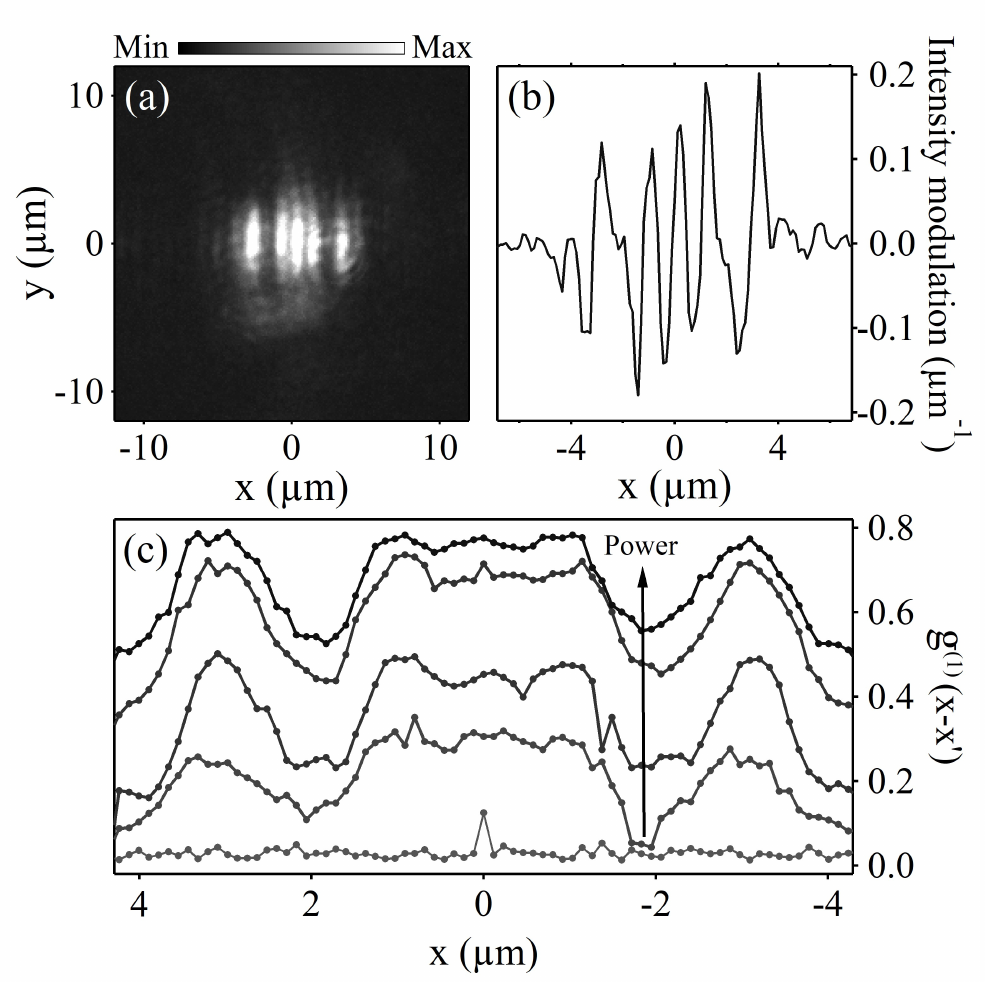}
		\caption{\textbf{{Formation of the moir\'e exciton polariton supersolid.}} (a) Real space image of the moir\'e exciton polariton supersolid. (b) Density modulation amplitude of the moir\'e polariton supersolid. (c) First-order spatial coherence functions at different pumping densities (from bottom to top: 0.3 P$_{th}$, 1.2 P$_\textup{th}$, 1.3 P$_\textup{th}$, 1.4 P$_\textup{th}$, 1.5 P$_\textup{th}$, here P$_\textup{th}$= 10.5 $\mu$J/cm$^2$).} 
	\end{figure}
	
	With the onset of parametric scattering under the large pumping density, the supersolid phase and density modulation can be triggered \cite{polariton supersolid2, polariton supersolid1, OPO theory1, OPO theory2}. The usual supersolids show a uniform periodic fringe due to the breaking of the translational symmetry within the system. In our experiment, the real space image of the perovskite CsPbBr$_3$ microplate shows a uniform distribution for polaritons when the pumping density is smaller than 10.5 $\mu$J/cm$^2$ (shown in SM), which transforms into a moir\'e pattern that can be decomposed of two one-dimensional periodic fringes with different periods along the same direction (Figure 3(a)) at the pumping density of 11.5 $\mu$J/cm$^2$. We note that the polariton distribution does not show any structures when condensation only occurs at $\ket{+k_0}$ state, which confirms the spontaneity of the density modulation in the parametric scattering process. This also excludes other artificial effects, like the confinement of the potential or pumping laser spot shape \cite{CP Lagoudakis,NP Bloch}. From the calculated derivative of the intensity lineprofile of the polariton condensate along \textit{x} direction, the density modulation is clearly observed, as shown in Figure 3(b). 
	
	
	This kind of polariton condensate with the moir\'e pattern indicates the existence of different kinds of supersolids from the cold atom condensates or BIC exciton polaritons \cite{polariton supersolid2, polariton supersolid1} in our microcavity. The long-range order of the superfluid part of the polariton condensate can be revealed by the experimental interferometric measurement, which is obtained by the first-order spatial coherence function $g^{(1)}$($x-x^{\prime}$) where $x$ and $x^{\prime}$ are the two positions within the polariton condensate. During the measurement, the mirror in one arm is replaced with a retro-reflector to reverse the real space image. We can see the transition from short-range correlation with nonzero value at zero time and distance of $g^{(1)}$($x-x^{\prime}$) below threshold to increased and extended coherence length above threshold (Figure 3(c)). The intensity modulation of the spatial coherence function is in phase with the density fringes, where the smaller spatial coherence positions coincide with the low-intensity region of the fringe within the polariton condensate. Especially enhanced spatial coherence is observed across the polariton condensate, which confirms the overall spatial coherence is built \cite{polariton supersolid1, polariton supersolid2}. From this, our supersolid is different from the fragmented polariton condensate or other isolated condensate arrays where the spatial or temporal coherence is greatly reduced. More importantly, the superfluidity of the polariton supersolids can be confirmed by the disappearance of the defect scattering which is introduced onto the perovskite with the femtosecond laser \cite{OPO Baowei}, as shown in the SM.

	\begin{figure}
		\centering
		\includegraphics[width=1\linewidth]{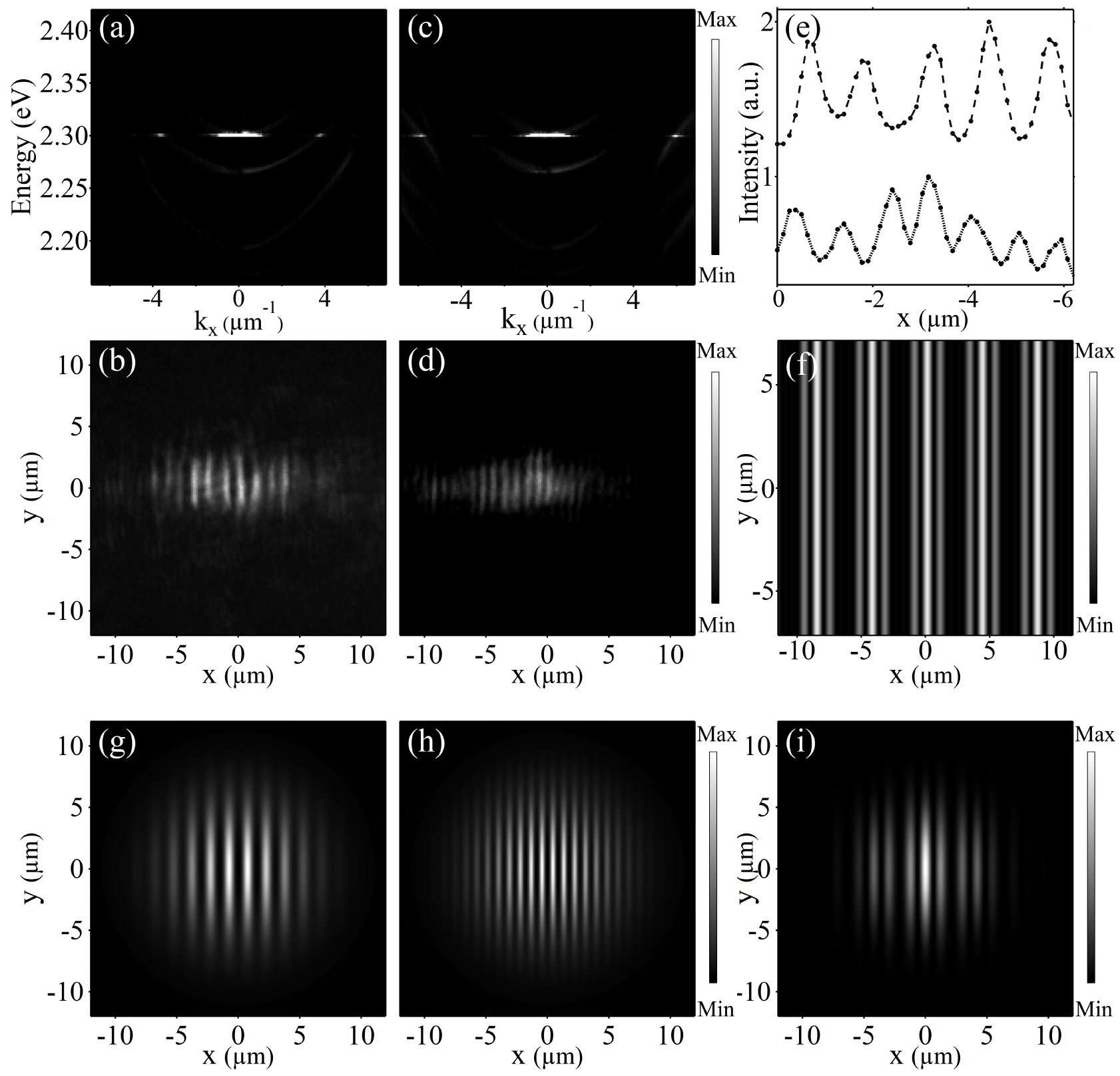}
		\caption{\textbf{{Two constituent polariton supersolids and simulated moir\'e state.}} (a) Dispersion taken for state $\ket{k_0}$ within LP0 and $\ket{\pm{k_1}}$ within LP1. (b) Real space image corresponding to (a). (c) Dispersion taken for state $\ket{k_0}$ with LP0 and $\ket{\pm{k_2}}$ within LP2. (d) Real space image corresponding to (c). (e) Lineprofiles of the density modulation of the two supersolids. (f) Calculated moir\'e exciton polariton supersolid pattern. (g, h) Simulated polariton supersolids corresponding to (b, d). (i) Simulated moir\'e polariton supersolid. (g-i) are obtained from GP equations. }
	\end{figure}
	
	Above moir\'e exciton polariton supersolid is formed due to the superposition of two constituent supersolids which are created at the same time. Two parametric scattering processes shown in Figure 2(e) produce two supersolids with different density modulation periods which are determined by the wavevectors ($k_1$, -$k_1$) of the states $\ket{\pm{k_1}}$ within LP1 and the wavevectors ($k_2$, -$k_2$) of the states $\ket{\pm{k_2}}$ within LP2. We measure the real space images when a filter is used in the momentum space (the filtered dispersion is plotted in Figure 4(a)) to extract the modes at the ground states $\ket{k_0}$ of LP0 and $\ket{\pm{k_1}}$ of LP1, which is shown in Figure 4(b). Here, a periodic fringe is observed with the period of 1.40 $\mu$m (the lineprofile is shown in Figure 4(e), the dashed one), which is consistent with the wavevector difference between these modes (2$\pi$/$k_1$=1.53 $\mu$m). It is worth noting that the linear polarizations of LP0 and LP1 are not perfectly orthogonal against each other due to the imperfection of the LC molecule orientation alignment in our microcavity (Figure 2(a) and Figure 2(b)), so there is still a small linear coupling between polariton branch LP0 and LP1. In this case, the phase for the periodic fringes within the supersolids is fixed \cite{polariton supersolid1, polariton supersolid2}. In this context, we can observe clear fringes accompanying the formation of the supersolid under the excitation of our femtosecond laser with the repetition rate of 1000 Hz. If the ground states $\ket{k_0}$ of LP0 and the modes $\ket{\pm{k_2}}$ of LP2 are extracted (see Figure 4(c)), we observe another periodic fringe with a period of 0.90 $\mu$m (Figure 4(d), the lineprofile is presented in Figure 4(e), the dotted one), which is also consistent with the wavevector difference between the two modes selected (2$\pi$/$k_2$=1.03 $\mu$m). This means that smaller wavevector difference results in a large density modulation period, and it clearly shows the existence of two supersolids in the double parametric scattering process within the LC microcavity. 
	
	\begin{figure}
		\centering
		\includegraphics[width=1\linewidth]{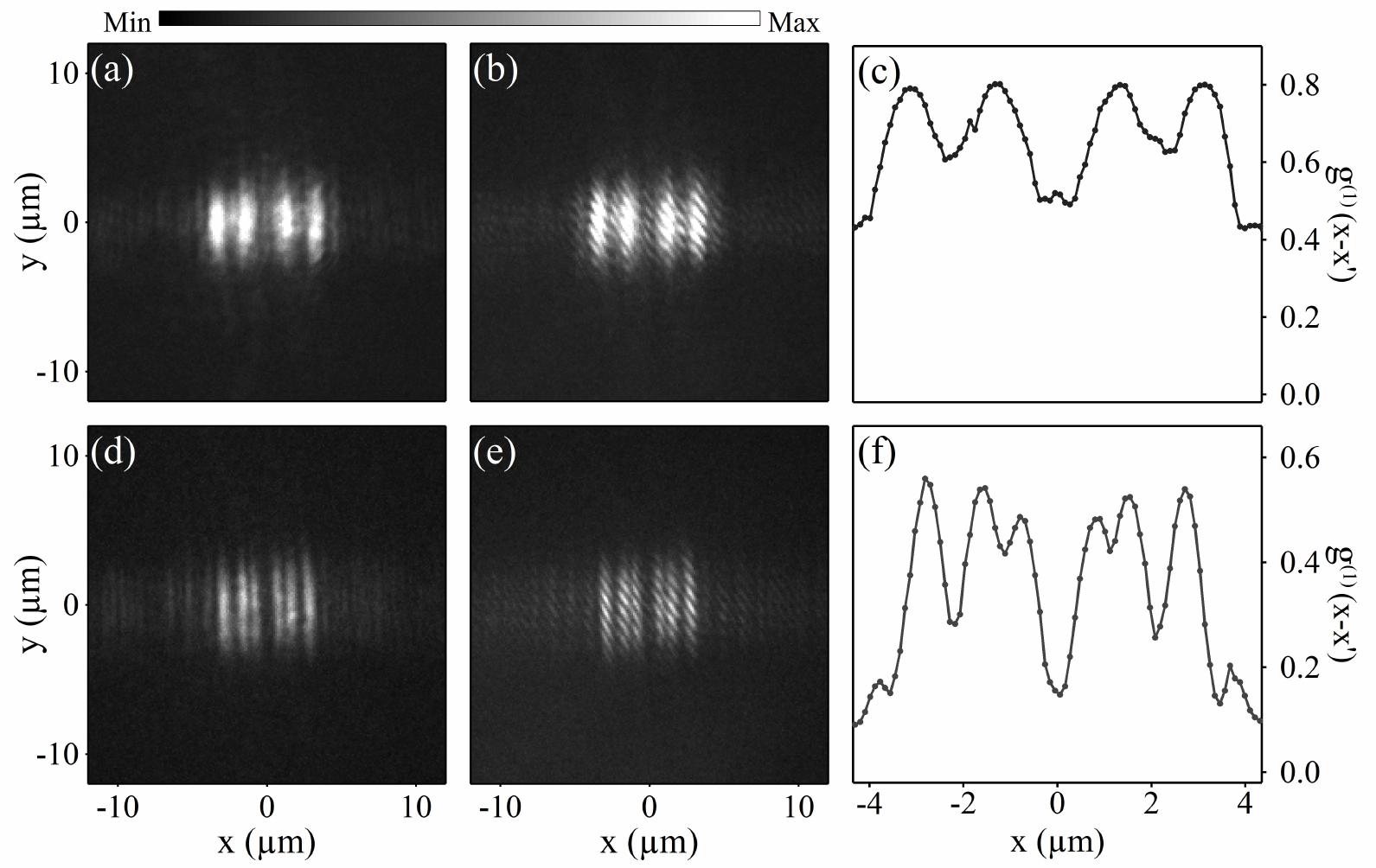}
		\caption{\textbf{{Nonrigidity of the moir\'e exciton polariton supersolids.}} (a, d) Moir\'e exciton polariton supersolids under the voltage of 1.7 V and 2 V, respectively. (b, e) Interference of the moir\'e exciton polariton supersolids of (a, d). (c, f) First-order spatial correlation functions of (a) and (d). The pumping density is the same as in Figure 3(a) and Figure 4. }
	\end{figure}
	
	The periods of the two supersolids are nearly commensurable thus we can observe a long period of 4.2 $\mu$m, and our microcavity can also resolve some fine structure of the moir\'e exciton polariton supersolid (three main stripes). Other fine structures of the moir\'e stripe cannot be resolved due to the limited resolution in our optical spectroscopy setup. To theoretically reproduce the moir\'e exciton polariton supersolid, we use two periodic fringes with the same intensity distribution as in Figure 4(b) and Figure 4(d). The superimposed total image (Figure 4(f)) agrees with the experimental results in Figure 3(a). The characteristic moir\'e stripe patterns for the supersolids are also evidenced by theoretical calculations, where the optical parametric oscillation \cite{PhysRevB.62.R4825} is taken into consideration in the Gross-Pitaevskii equation of polaritons under the non-resonant pumping (Figure 4(g-i), superfluidity in the supersolid phase is also confirmed, details in SM).

	The period of the density modulation fringes of the supersolid is determined by the difference of the wavevectors of the pump $k_0$, signal $k_1$ ($k_2$) and idler -$k_1$ (-$k_2$) modes in the parametric scattering process. In our LC microcavity, polaritons always condense at the ground state $\ket{k_0}$ of LP0 with the wavevector of $k_0$= 0 $\mu$m$^{-1}$ which is fixed under different voltages. When the voltage is increased from 1.6 V to 1.7 V, the LP1 is tuned towards higher energy due to smaller effective refractive index along the horizontal direction. Thus the wavevectors of the signal (idler) modes ($k_1$, -$k_1$) within this branch are decreased \cite{1-liquid crystal_science, Yao RD, Xiaokun vortex} (the relation with the applied voltage is plotted in the SM). In this context, the wavevector difference between the pump and signal (idler) states is reduced, which leads to a large period of the 1D density modulation fringes within the supersolid. On the other hand, the branch LP2 is not sensitive to the voltage onto the microcavity due to the vertical linear polarization, so the wavevectors of the signal and idler states within the branch LP2 ($k_2$, -$k_2$) are constant, and the polariton density modulation fringes created between LP0 and LP2 are fixed. In this case, the moir\'e pattern superimposed by these two 1D supersolids is electrically tuned into different shapes with different periods, as plotted in Figure 5(a). We check the existence of the long-range spatial coherence under 1.7 V by measuring the first-order spatial correlation functions, as plotted in Figure 5(b) and Figure 5(c). Continuously increasing the voltage to 2 V can tune the moir\'e polariton supersolid further in the real space, as shown in Figure 5(d-f). 
	
	A movie of the electrically tunable moir\'e polariton supersolids is shown in the SM where the period of the moir\'e polariton supersolids as the function of the voltage is included. In addition, time dependent moir\'e polariton supersolid dynamics are also plotted, which share some kind of similarity with the compression mode in the cold atom supersolids \cite{nonrigid 1, nonrigid 2}. Above results clearly reveal the nonrigidity of the moir\'e polariton supersolid, modulated by the electric method at room temperature. It is worth noting that our experimental scheme is realized with four finite-momentum states, the roton modes at finite-momentum could be explored by manipulating the wavevectors or momentum of the polariton condensations in the future.

	To summarize, we create a new spontaneously formed moir\'e state by simultaneously exciting two supersolids without the need of fabricating twisted complex periodic photonic structures \cite{photonic moire lattice} or graphene/TMD doublelayers \cite{flat band bilayer1, moire polariton} where the periodicity originates from the designed structure parameters. The density modulation fringes within our moir\'e polariton supersolids result from the spontaneous breaking of U(1) symmetry and translational symmetry in the degenerate parametric scattering process. The moir\'e polariton supersolid can be electrically tuned by modulating the wavevectors and the period of one constituent supersolid, revealing the nonrigidity of the supersolid. Compared with cold atom supersolid, our work shows a simple way to realize a new moir\'e quantum matter with two out-of-equilibrium supersolids (more in the future), which provides to study moir\'e induced nontrivial physics like flat/topological bands within the supersolid phase at room temperature by electrical means.

	\begin{acknowledgments}
		T.Gao acknowledges the support from the National Natural Science Foundation of China (NSFC, No. $12174285$, $12474315$). X. Zhai acknowledges the support from the National Natural Science Foundation of China(NSFC, No. 12504372) and the China Postdoctoral Science Foundation-Tianjin Joint Support Program under Grant Number (No. 2025T003TJ). 
	\end{acknowledgments}

	
	


\begin{thebibliography}{99}
		
		\bibitem{cold atom BEC} M. H. Anderson, J. R. Ensher, M. R. Matthews, C. E. Wieman and E. A, Cornell, Observation of Bose-Einstein Condensation in a Dilute Atomic Vapor. \textit{Science} \textbf{269,} 5221 (1995).
		
		
		
		
		
		\bibitem{BEC superfluid} N. Defenu, T. Donner, T. Macr{\`i}, G. Pagano, S. Ruffo, and A.
		Trombettoni. \textit{Review of Modern Physics} \textbf{95,} 035002 (2023).
		
		
		
		\bibitem{supersolid proposal 1} E. P. Gross, Unified theory of interacting bosons. \textit{Physical Review} \textbf{106,} 161 (1957).
		
		\bibitem{supersolid proposal 2} A. F. Andreev and I. M. Lifshitz, Quantum theory of defects in
		crystals. \textit{Soviet Physics JETP} \textbf{29,} 1107 (1969).
		
		\bibitem{supersolid proposal 3} G. V. Chester, Speculations on Bose-Einstein condensation
		and quantum crystals. \textit{Physical Review A} \textbf{2,} 256 (1970).
		
		\bibitem{supersolid proposal 4}  A. J. Leggett, Can a solid be “superfluid”? \textit{Physical Review Letters} \textbf{25,} 1543 (1970).
		
		
		\bibitem{helium} M. Boninsegni and N. V. Prokof’ev, Colloquium: Supersolids:
		What and where are they? \textit{Reviews of Modern Physics} \textbf{84,} 759 (2012).
		
		\bibitem{supersolid dipolar 1} L. Tanzi, E. Lucioni, F. Fam{\`a}, J. Catani, A. Fioretti, C. Gabbanini, R. N. Bisset, L. Santos, and G. Modugno, Observation of a dipolar quantum gas with metastable supersolid properties. \textit{Physical Review Letters} \textbf{122,} 130405 (2019).
		
		\bibitem{supersolid dipolar 2} F. B{\"o}ttcher, J.-N. Schmidt, M. Wenzel, J. Hertkorn, M. Guo, T.
		Langen, and T. Pfau, Transient Supersolid Properties in an Array of Dipolar Quantum Droplets. \textit{Physical Review X} \textbf{9,} 011051 (2019).
		
		\bibitem{supersolid dipolar 3} L. Chomaz, D. Petter, P. Ilzh{\"o}fer, G. Natale, A. Trautmann, C. Politi, G. Durastante, R. M. W. van Bijnen, A. Patscheider, M. Sohmen, M. J. Mark, F. Ferlaino, Long-Lived and transient supersolid behaviors in dipolar quantum gases. \textit{Physical Review X} \textbf{9,} 021012 (2019).
		
		\bibitem{supersolid dipolar 4} M. A. Norcia, C. Politi, L. Klaus, E. Poli, M. Sohmen, M. J. Mark, R. N. Bisset, L. Santos, F. Ferlaino, Two-dimensional supersolidity in a dipolar quantum gas. \textit{Nature} \textbf{596,} 357–361 (2021).
		
		\bibitem{supersolid dipolar 5} M. Guo, F. B{\"o}ttcher, J. Hertkorn, J.-N. Schmidt, M. Wenzel, H. P. Büchler, T. Langen, T. Pfau, The low-energy Goldstone mode in a trapped dipolar supersolid. \textit{Nature} \textbf{564,} 386–389 (2019).
		
		\bibitem{supersolid SOI 1} J.-R. Li, J. Lee, W. Huang, S. Burchesky, B. Shteynas, F. \c{c}.
		Top, A. O. Jamison, and W. Ketterle, A stripe phase with supersolid properties in spin-orbit-coupled Bose Einstein condensates. \textit{Nature} \textbf{543,} 91 (2017).
		
		\bibitem{supersolid SOI 2} A. Putra, F. Salces-C{\' a}rcoba, Y. Yue, S. Sugawa, and I. B.
		Spielman, Spatial Coherence of Spin-Orbit-Coupled Bose Gases. \textit{Physical Review Letters} \textbf{124,} 053605 (2020).
		
		\bibitem{supersolid cavity 1}J. L{\' e}onard, A. Morales, P. Zupancic, T. Donner, T. Esslinger,  Monitoring and manipulating Higgs and Goldstone modes in a supersolid quantum gas. \textit{Science} \textbf{358,} 1415–1418 (2017).
		
		\bibitem{supersolid cavity 2} J. L{\' e}onard, A. Morales, P. Zupancic, T. Esslinger, and T.
		Donner, Supersolid formation in a quantum gas breaking a continuous translational symmetry. \textit{Nature} \textbf{543,} 87 (2017).
		
		\bibitem{nonrigid 1} K. T. Geier, G. I. Martone, P. Hauke, W. Ketterle, S. Stringari, Dynamics of Stripe Patterns in Supersolid Spin-Orbit-Coupled Bose Gases. \textit{Physical Review Letters} \textbf{130,} 156001 (2023).
		
		
		\bibitem{nonrigid 2}  C. S. Chisholm, S. Hirthe, V. B. Makhalov, R. Ramos, R. Vatr{\' e}, J. Cabedo, A. Celi, and L. Tarruell, Probing supersolidity through excitations in a spin-orbit-coupled Bose-Einstein condensate. (2024), arXiv:2412.13861v1 [cond-mat.quant-gas]. 
		
		\bibitem{microcavity book} A. Kavokin, J. J. Baumberg, G. Malpuech, and F. P. Laussy, Microcavities. Series on Semiconductor Science and Technology (2007).
		
		\bibitem{polariton supersolid1} D. Trypogeorgos, A. Gianfrate, M. Landini, D. Nigro, D. Gerace, I. Carusotto, F. Riminucci, K. W. Baldwin, L. N. Pfeiffer, G. I. Martone, M. De Giorgi, D. Ballarini, D. Sanvitto, Emerging supersolidity from a polariton condensate in a photonic crystal waveguide. \textit{Nature} \textbf{639,} 337-341 (2025).
		
		\bibitem{polariton supersolid2} D. Nigro, D. Trypogeorgos, A. Gianfrate, D. Sanvitto, I. Carusotto, D. Gerace, Supersolidity of Polariton Condensates in Photonic Crystal Waveguides. \textit{Physical Review Letters} \textbf{134,} 5 (2025).
		
		\bibitem{OPO1} F. Chen, H. Zhou, H. Li, J. Cao, S. Luo, Z. Sun, Z. Zhang, Z. Shao, F. Sun, B. Zhou, H. Dong, H. Xu, H. Xu, A. Kavokin,
		Z. Chen, and J. Wu,  Femtosecond Dynamics of a Polariton Bosonic Cascade at Room Temperature. \textit{Nano Letters} \textbf{22,} 2023 (2022).
		
		\bibitem{OPO2} Z. Ye, F. Chen, H. Zhou, S. Luo, Z. Sun, H. Xu, H. Xu, H. Li,
		Z. Chen, and J. Wu, Ultrafast intermode parametric scattering dynamics in room-temperature polariton condensates. \textit{Physical Review B} \textbf{107,} L060303 (2023).
		
		\bibitem{OPO3} W. Xie, H. Dong, S. Zhang, L. Sun, W. Zhou, Y. Ling, J. Lu,
		X. Shen, and Z. Chen, Room- temperature Polariton Parametric Scattering Driven by a One-dimensional Polariton Condensate. \textit{Physical Review Letters} \textbf{108,} 166401 (2012).
		
		\bibitem{PRL1} C. Leyder, T. C. H. Liew, A. V. Kavokin, I. A. Shelykh, M. Romanelli, J. P. Karr, E. Giacobino, A. Bramati, Interference of coherent polariton beams in microcavities: polarization-controlled optical gates. \textit{Physical Review Letters} \textbf{99,} 196402 (2007).
		\bibitem{Nanophotonics1} K. Sawicki, T. J. Sturges, M. {\' S}ciesiek, T. Kazimierczuk, K. Sobczak, A. Golnik, W. Pacuski, J. Suffczy{\' n}ski, Polariton lasing and energy-degenerate parametric scattering in non-resonantly driven coupled planar microcavities. \textit{Nanophotonics} \textbf{10,} 2421-2429 (2021).
		
		\bibitem{1-liquid crystal_science} K. Rechci{\' n}ska, M. Kr{\' o}l, R. Mazur, P. Morawiak, R. Mirek, K. \L{}empicka, W. Bardyszewski, M. Matuszewski, P. Kula, W. Piecek, P. G. Lagoudakis, B. Pi\c{e}tka, J. Szczytko, Engineering spin-orbit synthetic Hamiltonians in liquid-crystal optical cavities. \textit{Science} \textbf{366,} 727-730 (2019).
		
		\bibitem{stripe phase} M. Muszy{\' n}ski, P. Kokhanchik, D. Urbonas, P. Kapu{\' s}ci{\' n}ski, P. Oliwa, R. Mirek, I. Georgakilas, T. St{\"o}ferle, R. F. Mahrt, M. Forster, U. Scherf, D. Dovzhenko, R. Mazur, P. Morawiak, W. Piecek, P. Kula, B. Pi\c{e}tka, D. Solnyshkov, G. Malpuech, J. Szczytko, Observation of a stripe phase in a spin-orbit coupled exciton-polariton Bose-Einstein condensate. (2024), arXiv:2407.02406v1 [cond-mat.mes-hall]. 
		
		\bibitem{Yao RD} Y. Li, X. Ma, X. Zhai, M. Gao, H. Dai, S. Schumacher, T. Gao, Manipulating polariton condensates by Rashba-Dresselhaus coupling at room temperature.  \textit{Nature Communications} \textbf{13,} 3785 (2022).
		
		\bibitem{Xiaokun vortex} X. Zhai, X. Ma, Y. Gao, C. Xing, M. Gao, H. Dai, X. Wang, A. Pan, S. Schumacher, T. Gao, Electrically controlling vortices in a neutral exciton-polariton condensate at room temperature. \textit{Physical Review Letters} \textbf{131,} 136901 (2023).
		
		\bibitem{gao ying} Y. Gao, X. Ma, X. Zhai, C. Xing, M. Gao, H. Dai, H. Wu, T. Liu, Y. Ren, X. Wang, A. Pan, W. Hu, S. Schumacher, and T. Gao, Single-shot spatial instability and electric control of polariton condensates at room temperature. \textit{Physical Review B}  \textbf{108}, 205303(2023).
		
		\bibitem{zhuxiaoyang} M. S. Spencer, Y. Fu,  A. P. Schlaus, D. Hwang,  Y. Dai,  M. D. Smith, D. R. Gamelin, X.-Y.  Zhu, Spin-orbit coupled exciton-polariton condensates in lead halide perovskites. \textit{Science Advances} \textbf{7,} eabj7667 (2021).
		
		\bibitem{21-xiong condensation} R. Su, J. Wang, J. Zhao, J. Xing, W. Zhao, C. Diederichs, T. C. H. Liew, Q. Xiong, Room temperature long-range coherent exciton polariton condensate flow in lead halide perovskites. \textit{Science Advances} \textbf{4,} eaau0244 (2018).
		
		
		\bibitem{nonequilibrium} M. Wouters, I. Carusotto, C. Ciuti, Spatial and spectral shape of inhomogeneous nonequilibrium exciton-polariton condensates. \textit{Physical Review B} \textbf{77,} 115340 (2008).
		
		\bibitem{polarization OPO} G. Dasbach, C. Diederichs, J. Tignon, C. Ciuti, P. Roussignol, C. Delalande, M. Bayer, A. Forchel, Polarization inversion via parametric scattering in quasi-one-dimensional microcavities. \textit{Physical Review B}  \textbf{71,} 161308 (2005).
		
		\bibitem{OPO theory1} M. Wouters, I. Carusotto, Goldstone mode of optical parametric oscillators in planar semiconductor microcavities
		in the strong-coupling regime. \textit{Physical Review A} \textbf{76,} 043807 (2007). 
		
		\bibitem{OPO theory2} I. Carusotto, C. Ciuti, Spontaneous microcavity-polariton coherence across the parametric threshold: Quantum Monte Carlo studies. \textit{Physical Review B}  \textbf{72,} 125335 (2005).
		
		\bibitem{CP Lagoudakis} J. D. T{\"o}pfer, H. Sigurdsson, L. Pickup, P. G. Lagoudakis, Time-delay polaritonics. \textit{Communications Physics}  \textbf{3,} 2 (2020).
		
		\bibitem{NP Bloch} E. Wertz, L. Ferrier, D. D. Solnyshkov, R. Johne, D. Sanvitto, A. Lema\^{i}tre, I. Sagnes, R. Grousson, A. V. Kavokin, P. Senellart, G. Malpuech, J. Bloch, Spontaneous formation and optical manipulation of extended polariton condensates. \textit{Nature Physics}  \textbf{6,} 860-864 (2010).
		
		\bibitem{OPO Baowei} K. Peng, R. Tao, L. Haeberl{\' e}, Q. Li, D. Jin, G. R. Fleming, S. K{\' e}na-Cohen, X. Zhang, W. Bao, Room-temperature polariton quantum fluids in halide perovskites. \textit{Nature Communications}  \textbf{13,} 7388 (2022).
		
		\bibitem{PhysRevB.62.R4825} C. Ciuti, P. Schwendimann, B. Deveaud, A. Quattropani, Theory of the angle-resonant polariton amplifier. \textit{Physical Review B}  \textbf{62,} R4825-R4828 (2000).
		
		
		\bibitem{photonic moire lattice} P. Wang, Y. Zheng, X. Chen, C. Huang, Y. V. Kartashov, L. Torner, V. V. Konotop, and F. Ye, Localization and delocalization of light in photonic moir{\'e} lattices. \textit{Nature} \textbf{577}, 42-46(2019). 
		
		\bibitem{flat band bilayer1} Y. Cao, V. Fatemi, A. Demir, S. Fang, S. L. Tomarken, J. Y. Luo, J. D. Sanchez-Yamagishi, K. Watanabe, T. Taniguchi, E. Kaxiras, R. C. Ashoori, and P. Jarillo-Herrero, Correlated insulator behaviour at half-filling in magic-angle graphene superlattices. \textit{Nature} \textbf{556}, 80 (2018). 
		
		
		\bibitem{moire polariton} L. Zhang, F. Wu, S. Hou, Z. Zhang, Y. H. Chou, K. Watanabe, T. Taniguchi, S. R. Forrest, and H. Deng, Van der Waals heterostructure polaritons with moir{\'e}-induced nonlinearity. \textit{Nature} \textbf{591}, 61–65 (2021). 
		
		
	\end{thebibliography}
\end{document}